\documentclass[12pt]{article}

\usepackage{natbib}
\usepackage[table]{xcolor}
\usepackage{lmodern}
\usepackage{graphicx}
\usepackage{subcaption}
\usepackage{amsmath}
\usepackage{dsfont}
\usepackage{placeins}
\usepackage{multirow}
\usepackage{tabularray}
\usepackage{hyperref}
\usepackage{enumerate}

\usepackage{algpseudocode}
\usepackage{amsthm}
\usepackage{hyperref}
\usepackage{mathrsfs}
\usepackage{amsfonts}
\usepackage{bbding}
\usepackage{listings}
\usepackage{appendix}
\usepackage{fancyvrb}
\usepackage{cancel}
\usepackage{algorithm}

\newcounter{probnum}
\allowdisplaybreaks

\definecolor{tabblue}{rgb}{.870588,.905882,.94902}
\definecolor{gray}{rgb}{0.7,0.7,0.7}
\definecolor{black}{rgb}{0,0,0}
\definecolor{white}{rgb}{1,1,1}
\definecolor{blue}{rgb}{0.0,0.0,1}

\definecolor{green}{rgb}{0,0.5,0}

\definecolor{yellow}{rgb}{1,0.549,0}

\definecolor{red}{rgb}{0.6,0.0,0.0}

\definecolor{darkred}{rgb}{0.9,0.4,0}

\definecolor{purple}{rgb}{0.58,0,0.827}

\definecolor{backgcode}{rgb}{0.97,0.97,0.8}
\definecolor{Brown}{cmyk}{0,0.81,1,0.60}
\definecolor{OliveGreen}{cmyk}{0.64,0,0.95,0.40}
\definecolor{CadetBlue}{cmyk}{0.62,0.57,0.23,0}

\newcommand{\qu}[1]{``{#1}''}
\newcommand{\bv}[1]{\boldsymbol{#1}}

\newcommand{\bSigmaw}{\bv{\Sigma}_{\W}}

\newcommand{\bSigmawB}{\bv{\Sigma}_{\W_B}}

\newcommand{\inddist}{~{\buildrel ind \over \sim}~}

\newcommand{\half}{\frac{1}{2}}

\newcommand{\A}{\bv{A}}

\newcommand{\Z}{\bv{Z}}

\newcommand{\p}{\bv{p}}

\newcommand{\Y}{\bv{Y}}
\newcommand{\W}{\bv{W}}

\newcommand{\x}{\bv{x}}

\newcommand{\w}{\bv{w}}

\newcommand{\onevec}{\bv{1}}

\newcommand{\y}{\bv{y}}

\newcommand{\btau}{\bv{\tau}}

\renewcommand{\v}{\bv{v}}

\newcommand{\reals}{\mathbb{R}}

\newcommand{\limitn}{\lim_{n \rightarrow \infty}}

\newcommand{\beqn}{\vspace{-0.25cm}\begin{eqnarray*}}
\newcommand{\eeqn}{\end{eqnarray*}}
\newcommand{\bneqn}{\vspace{-0.25cm}\begin{eqnarray}}
\newcommand{\eneqn}{\end{eqnarray}}
\newcommand{\benum}{\begin{enumerate}}
\newcommand{\eenum}{\end{enumerate}}

\newcommand{\parens}[1]{\left(#1\right)}
\newcommand{\squared}[1]{\parens{#1}^2}

\newcommand{\prob}[1]{\mathbb{P}\parens{#1}}

\newcommand{\cprob}[2]{\prob{#1~|~#2}}

\newcommand{\Lp}[1]{\mathbb{L}^{#1}}

\newcommand{\bracks}[1]{\left[#1\right]}
\newcommand{\braces}[1]{\left\{#1\right\}}

\newcommand{\abss}[1]{\left|#1\right|}
\newcommand{\norm}[1]{\left|\left|#1\right|\right|}
\newcommand{\normsq}[1]{\norm{#1}^2}

\newcommand{\expe}[1]{\mathbb{E}\bracks{#1}}

\newcommand{\cexpe}[2]{\expe{#1\,|\,#2}}

\newcommand{\cvar}[2]{\var{#1\,|\,#2}}

\newcommand{\var}[1]{\mathbb{V}\text{ar}\bracks{#1}}

\newcommand{\cov}[2]{\mathbb{C}\text{ov}\bracks{#1,\,#2}}

\newcommand{\oneover}[1]{\frac{1}{#1}}

\newcommand{\mathand}{~~\text{and}~~}

\newcommand{\normnot}[2]{\mathcal{N}\parens{#1,\,#2}}

\newcommand{\stdnormnot}{\normnot{0}{1}}

\newcommand{\convLp}[1]{~{\buildrel \Lp{#1} \over \rightarrow}~}

\newcommand{\convd}{~{\buildrel d \over \longrightarrow}~}

\usepackage{fancyhdr}
\usepackage{abstract}
\usepackage[auth-sc,affil-sl]{authblk}
\usepackage[margin=1.0in]{geometry}
\usepackage{amssymb}
\usepackage{algpseudocode}

\newtheorem{theorem}{Theorem}[section]
\newtheorem{remark}{Remark}[section]
\newtheorem{lemma}{Lemma}[theorem]

\newcommand{\diag}[1]{\text{diag}\bracks{#1}}
\newcommand{\tr}[1]{\text{tr}\bracks{#1}}

\newcommand{\ourtitle}{Block Designs that Provide Optimal Power in the Cochran–Mantel–Haenszel Test}
\title{\ourtitle}

\author[1]{David Azriel\thanks{Electronic address: \texttt{davidazr@technion.ac.il}}}
\author[2]{Adam Kapelner\thanks{Electronic address: \texttt{kapelner@qc.cuny.edu}; Principal Corresponding author}}
\author[3]{Abba M. Krieger\thanks{Electronic address: \texttt{krieger@wharton.upenn.edu}}}

\affil[1]{\small Faculty of Data and Decision Sciences, The Technion, Haifa, Israel}
\affil[2]{\small Department of Mathematics, Queens College, CUNY, USA}
\affil[3]{\small Department of Statistics, The Wharton School of the University of Pennsylvania, USA}

\begin{document}
\maketitle

\begin{abstract}
We consider the asymptotic power performance under local alternatives of the Cochran-Mantel-Haenszel test. Our setting is non-traditional: we investigate randomized experiments that assign subjects via Fisher's blocking design. We show that blocking designs that satisfy a certain balance condition are asymptotically optimal. When the potential outcomes can be ordered, the balance condition is met for all blocking designs with number of blocks going to infinity. More generally, we prove that the pairwise matching design of Greevy et al. (2004) satisfies the balance condition under mild assumptions. In smaller sample sizes, we show a second order effect becomes operational thereby making blocking designs with a smaller number optimal. In practical settings with many covariates, we recommend pairwise matching for its ability to approximate the balance condition.

\end{abstract}

\section{Background}\label{sec:intro}

We consider a classic problem: an experiment with $2n$ \emph{subjects} (\emph{individuals}, \emph{participants} or \emph{units}). The experiment has two \emph{arms} (\emph{treatments}, \emph{manipulations} or \emph{groups}), which we will call treatment and control. Subjects have $p$ observed subject-specific \emph{covariates} (\emph{measurements}, \emph{features} or \emph{characteristics}) denoted $\x_i \in \reals^p$ for the $i$th subject.  We consider one incidence (zero or one) \emph{outcome} (\emph{response} or \emph{endpoint}) of interest for the $2n$ subjects denoted $\y = \bracks{y_1, \ldots, y_{2n}}^\top$. The investigator then wishes to test if the true odds ratio across the two arms is one (or the log odds ratio is zero).

The setting we investigate is where all $\x_i$'s are known beforehand and considered fixed. This setting is common in industry and is also found in \qu{many phase I studies [that] use `banks' of healthy volunteers ... [and] ... in most cluster randomised trials, the clusters are identified before treatment is started} \citep[page 1440]{Senn2013}.

Formally defined, the \emph{randomization} (\emph{allocation} or an \emph{assignment}) is a vector $\w = \bracks{w_1, \ldots, w_{2n}}^\top$ whose entries indicate whether the subject is placed into the treatment arm (coded numerically as +1) or placed into the control arm (coded numerically as -1). After the sample is provided, the only control the experimenter has is the choice of the entries in $\w$. The process whereby the experimenter summons this assignment vector is termed an experimental \emph{design} (\emph{strategy}, \emph{algorithm}, \emph{method} or \emph{procedure}). Randomized experimental design is thus a generalized multivariate Bernoulli random variable that realizes $\w$'s. 

One typical design is the balanced complete randomization design (BCRD, \citealp[p. 1171]{Wu1981}) which has also been called the \qu{gold-standard}. This design has $\binom{2n}{n}$ allocations each that satisfy the equal-arm-allocation constraint that $\w^\top \onevec_{2n} = 0$ and each of these allocations is then chosen with equal probability. However, there are many unlucky $\w$'s from BCRD that result in large differences in the distribution of observed covariates between the two arms. The amount of covariate value heterogeneity between groups we term \emph{observed imbalance}. In the classic setup of a linear model with continuous response, this observed imbalance creates estimation bias from the perspective of $\w$. Reducing this observed imbalance involves removing $\w$'s from the full support of the distribution which yield these observed imbalances. Hence such designs are termed \qu{restricted designs}.

Restricted designs have a long literature once again starting with \citet[p. 251]{Fisher1925} who wrote \qu{it is still possible to eliminate much of the \ldots [observed imbalance], and so increase the accuracy of our [estimator], by laying restrictions on the order in which the strips are arranged}. Here, he introduced the \emph{blocking design}, a restricted design still popular today. Taking this design to the extreme are $n$ blocks of two subjects each (i.e., pairs of subjects) where one is randomly allocated to the treatment arm and the other to the control arm. This is termed the pairwise matching (PM) design \citep{Greevy2004}. 

In our previous work that investigated blocking designs for general response type \citep{Azriel2024}, we considered estimation of the sample average treatment effect defined as $\oneover{2n} \onevec_{2n}^\top(\y_T - \y_C)$. Although this estimand is a natural target of inference for most outcome types, in the incidence setting this estimand (called the \qu{mean probability difference} or the \emph{risk difference}) is awkward; it averages large differences near 50\% and small differences near 0 and 1. In this work, we consider the odds ratio estimate. We then study the asymptotic power of a standard test, the \citet{Cochran1954}-\citet{Mantel1959} Test (CMHT).

The CMHT is usually employed in an observational study for testing the response difference between the two levels of a binary variable of interest where the data is further stratified based on other variables which are not being tested explicitly. The assumed null hypothesis is: there is no difference in outcomes among subjects in the different levels of the binary variable of interest. The utility of the CMHT is that it can both provide higher power (by averaging estimates within the levels of the other variables) and can also help to avoid Simpson's paradox (when there is confounding on the other variables). 

Our perspective change herein (which we believe to be novel) is that we optimize the CMHT in randomized experiments (not in observational studies). During experimentation, one has control over allocating $\w$ and can thus directly control for the strata (which are fixed by an observed variable in an observational study) as individual blocks. In other words, the binary variable of interest is $\w$ and the other variables are rendered into the strata of the nominal block category.

We show herein that asymptotically optimal power for the CMHT is PM when considering the space of block designs (as the unconditional space is intractable). We define our setting formally in Section~\ref{sec:setup}, discuss our theoretical contribution in Section~\ref{sec:methods}, provide simulation evidence of our results in Section~\ref{sec:simulations} and conclude in Section~\ref{sec:discussion}.

\section{Problem Setup}\label{sec:setup}

Consider the potential outcome framework of \citet{Rubin2005} and let $\p_T \in [0,1]^{2n}$ and $\p_C \in [0,1]^{2n}$ be the unobserved probabilities of positive response under the treatment and control respectively such that for all $i=1, \ldots, 2n$,


\bneqn\label{eq:bernoullis}
Y_{T,i} \inddist {\rm Bernoulli}(p_{T,i}) ~~\mathand~~ Y_{C,i} \inddist {\rm Bernoulli}(p_{C,i}).
\eneqn

\noindent In vector form, the $2n$ random responses $\Y$ can then be expressed as

\bneqn\label{eq:potential_outcomes}
\Y = \half\Big(\Y_T + \Y_C + \diag{\W} (\Y_T - \Y_C)\Big)
\eneqn

\noindent under the Stable Unit Treatment Values Assumption. We then assume that the design $\W$ is independent of the pair $(\Y_T,\Y_C)$.

As discussed in the introduction: we limit our scope to blocking designs. Specifically, we consider homogeneous balanced-allocation blocks, denote $B$ as the number of blocks, $n_B := 2n / B$ as the number of subjects per block (which is assumed to be a whole number) and $\W_B$ as the block design with $B$ blocks. The experimental data by block $b$ will then be denoted as follows:

\begin{table}[ht]
\centering
\begin{tabular}{c|cc|c} 
& Treatment ($w_i = +1$) & Control ($w_i = -1$) & Total \\ \hline
Affected ($y_i = 1) $   & $n_{T,1_b}$ & $n_{C,1_b}$ & $n_{1_b}$\\
Unaffected  ($y_i = 0$) & $n_{T,0_b}$ & $n_{C,0_b}$ & $n_{0_b}$  \\ \hline
Total &$n_B/2$ & $n_B/2$ & $n_B$\\
\end{tabular}
\end{table}

\noindent The CMHT statistic in our blocking setting can then be expressed as

\begin{equation}\label{eq:CMHT_statistic}
MH := \frac{
\squared{\,\displaystyle\sum_{b=1}^B\left( n_{T,1_b} -\frac{n_{1_b}}{2}\right)}
}{
\displaystyle\oneover{4(n_B-1)}\sum_{b=1}^B n_{1_b} n_{0_b}
} = \frac{(\W_B^\top \Y)^2  }{ \Y^\top \bSigmawB \Y }
\end{equation}

\noindent where the simplification on the right hand side follows from algebraic calculations found in Appendix~\ref{sec:calc}. The term $\bSigmawB := \var{\W_B}$ is the $2n \times 2n$ variance-covariance matrix of the experimental design. Since we consider blocking designs, the form of $\bSigmawB$ is block diagonal consisting of $B$ blocks of size $n_B \times n_B$ where each block consists of entries with 1 on the diagonal and $-1/(n_B - 1)$ on the off-diagonal. 

This $MH$ statistic is known to have an asymptotic chi-squared distribution with one degree of freedom under the null \citep[Section 6.3.2]{Agresti2002}. For the tractability of results, we consider the following related quantity which is asymptotically distributed as a standard normal under the null:

\beqn
\widetilde{MH} := \pm\sqrt{MH} = \frac{\W_B^\top \Y  }{ \sqrt{ \Y^\top \bSigmawB \Y }} \convd \stdnormnot.
\eeqn

\section{Main result}\label{sec:methods}

We wish to optimize the power of the CMHT over all block designs $\W_{B}$ with B blocks. We now define the following useful quantities:

\bneqn 
\btau &:=& \bracks{\frac{p_{T,1}-p_{C,1}}{2}~~\frac{p_{T,2}-p_{C,2}}{2}~~ \ldots ~~ \frac{p_{T,2n}-p_{C,2n}}{2}}^\top, \nonumber\\
~~\bar{\tau}_{n} &:=& \frac{1}{2n}\sum_{i=1}^{2n}\tau_i, \nonumber\\
\v &:=& \bracks{\frac{p_{T,1}+p_{C,1}}{2}~~\frac{p_{T,2}+p_{C,2}}{2}~~ \ldots ~~ \frac{p_{T,2n}+p_{C,2n}}{2}}^\top ~~\mathand \label{eq:def_of_v_vec} \\
\eta_n &:=& \oneover{2n} \parens{\v^\top \bSigmawB \v + \frac{\p_T^\top (\onevec - \p_T)+\p_C^\top (\onevec - \p_C)}{2}}. \label{eq:def_of_eta_vec}
\eneqn

\noindent It can be shown that $\expe{\frac{1}{2n} \W_{B}^\top \Y } = \bar{\tau}_n$ and $\var{\frac{1}{2n} \W_{B}^\top \Y } =\frac{1}{2n} \eta_n$ \citep[see][Appendix A.1 and Appendix A.2 respectively]{Kapelner2023}. For block $b$, define 

\beqn
\sigma^2_b:=\frac{n_b}{n_b-1}\sum_{i \in I_b}(v_i-\bar{v}_b)^2 \mathand \bar{v}_b:=\frac{1}{n_B}\sum_{i\in I_b} v_i
\eeqn

\noindent where $I_b$ is the set of indexes of block $b$. This notation allows us to express the critical quadratic form term $\v^\top \bSigmawB \v$ in Equation~\ref{eq:def_of_eta_vec} equivalently as $\sum_{b=1}^B \sigma_b^2$. We also define 

\beqn
\sigma^2_{min}:=\min\{\sigma_1^2,\ldots,\sigma_B^2\} \mathand \sigma^2_{max}:=\max\{\sigma_1^2,\ldots,\sigma_B^2\}. 
\eeqn

\noindent 
We allow the number both the block size ($n_B$) and the number of blocks ($B$) to vary with $n$ and consider two types of asymptotics:

\begin{enumerate}[(i)]

    \item  {\em Small blocks asymptotic}, in which $n_B$ is of smaller order than $B$, i.e., $\frac{n_B}{B} \to 0$, or equivalently, $\frac{\sqrt{n}}{B} \to 0$.
    
    \item {\em Large blocks asymptotic}, in which $n_B$ is of larger order than $B$, i.e., $\frac{B}{n_B} \to 0$, or equivalently, $\frac{B}{\sqrt{n}} \to 0$. In this regime we need an additional assumption, namely that $\sigma^2_{min}/\sigma^2_{max}$ is bounded. 
\end{enumerate}

In the literature of the CMHT two types of asymptotics are usually considered: fixed $B$ \citep[e.g.,][]{Zelen1971}, or fixed $n_B$ (e.g., \citealp{Breslow1981} and \citealp[Section 6.3.4]{Agresti2002}). We allow a more flexible regime in which both $n_B$ and $B$ go to infinity. Notice, however, that the case where $n_B$ and $B$ are of the same order is not covered by our asymptotics. For large blocks asymptotics we need an additional assumption, that $\sigma^2_{min}/\sigma^2_{max}$ is bounded. It is satisfied when the variances of the $v_i$'s among different blocks are comparable. It trivially holds when $B$ is fixed and $\v$ is nondegenerate. Our main result is now given.







\begin{theorem}\label{thm:MH_tilde_asymptotic_distr}
Consider the model and assumptions in Section \ref{sec:setup}. Consider a sequence of local alternative hypotheses where $\lim_{n \rightarrow \infty} \sqrt{2n} \bar{\tau}_{n} = c$. Assume $\frac{\p_T^\top (\onevec - \p_T)+\p_C^\top (\onevec - \p_C)}{2n}$ is bounded away from zero and assume ${\displaystyle \lim_{n \rightarrow \infty}} \frac{1}{2n}\normsq{\btau}=0$ and $\displaystyle\lim_{n \rightarrow \infty} \eta_n = \eta_{\infty} > 0$. 
Then,

\bneqn\label{eq:main_convergence_result}
\widetilde{MH} \convd \normnot{\frac{c}{\sqrt{\eta_{\infty}}}}{1}
\eneqn

\noindent in either the (i) small blocks asymptotic regime or the (ii) large blocks asymptotic regime. 
\end{theorem}

\noindent The proof of the theorem is given in Appendix \ref{sec:proof} and it involves non-standard asymptotics for two reasons. First, as mentioned previously we allow both $n_B$ and $B$ to go to infinity and second, $\W_B$ has dependent entries due to the design. To overcome these difficulties we use Berry-Essen-type bounds and results from Stein's approximation \citep{Chen2010}.

 Higher power means that the value of the asymptotic mean of $\widetilde{MH}$ should be made as large as possible; hence the value of $\eta_\infty$ should be made as small as possible over all considered designs. Recall the definition of $\eta_n$ (Equation~\ref{eq:def_of_eta_vec}), which is a sum of two terms. The second term does not depend on $\W_B$. It follows that blocking designs that send $\oneover{2n}\v^\top \bSigmawB$ to zero in turn maximize the asymptotic power. To interpret this condition notice that
 
\bneqn\label{eq:average}
\oneover{2n}\v^\top \bSigmawB \v =\frac{1}{B}\sum_{b=1}^B \frac{1}{n_B-1} \sum_{i \in I_b}(v_i-\bar{v}_b)^2.
\eneqn

\noindent 
This expression is the average variance of the $v_i$'s within each block. 
If the order of the $v_i$'s is known (see the discussion in Section \ref{subsec:simulation_p_1}) then the blocking procedure that minimizes $\oneover{2n}\v^\top \bSigmawB$ is to have the smallest $n_B$ $v_i$'s in the first block, the second smallest $n_B$ $v_i$'s in the second block etc; see  Section \ref{subsec:simulation_p_1}. When the blocking is done in this way, it is easy to see via Equation~\ref{eq:average} that $\oneover{2n}\v^\top \bSigmawB \to 0$ for any design with $B\to \infty$. More generally, we showed in \citet[Section A.6]{Kapelner2025} that $\oneover{2n}\v^\top \bSigmawB \to 0$ holds for the optimal nonbipartite pairwise matching algorithm when the covariates are bounded and $v_i = f(\x_i)$, where $f$ is Lipschitz continuous. 

However, there is finite sample behavior that is relevant to optimal design. In Appendix \ref{sec:proof}, we prove the following results: 

\bneqn
\oneover{2n} \Y^\top \bSigmawB \Y &\convLp{2}& \eta_\infty, \nonumber \\
\expe{ \frac{1}{2n}  \Y^\top \bSigmawB \Y}
&=& \frac{1}{2n} \parens{\v^\top \bSigmawB \v + \sum_{i=1}^{2n} v_i(1-v_i) + \frac{1}{n_B-1} \parens{ \normsq{\btau} - {\boldsymbol \tau}^\top \bSigmawB {\boldsymbol \tau}}} \nonumber 
\eneqn

\noindent and an easy calculation shows that

\bneqn\label{eq:small_n_effect_eqs}
\normsq{\btau} - {\boldsymbol \tau}^\top \bSigmawB {\boldsymbol \tau} = \frac{1}{n_B-1}\sum_{b=1}^B \sum_{i_1,i_2 \in I_B;i_1\ne i_2} \tau_{i_1} \tau_{i_2}.
\eneqn

\noindent The quantity on the left hand side of Equation~\ref{eq:small_n_effect_eqs} is positive if $\tau_i$ has the same sign for all $i$'s in the same block. If this is the case, $\frac{1}{n_B-1} \left( \normsq{\btau} - {\boldsymbol \tau}^\top \bSigmawB {\boldsymbol \tau}   \right)$ is large when $n_B$ is small. This term vanishes asymptotically, but for small $n$ it penalizes designs with small block size by shifting the mean in Equation~\ref{eq:main_convergence_result} towards zero. This finding is demonstrated in Section \ref{subsec:simulation_p_1}. We summarize the discussion of this section in our flagship result below.
\begin{remark}\label{rem:pm_optimal}

Under the assumptions of Theorem \ref{thm:MH_tilde_asymptotic_distr}, every design that satisfies 

\beqn
\limitn \oneover{2n}\v^\top \bSigmawB \v = 0
\eeqn

\noindent maximizes the asymptotic power. 
When the order of the $v_i$'s is known, the condition is met for blocking design that sends $B\to \infty$. If the order is not known, the nonbipartite pairwise matching algorithm satisfies the condition (under mild assumptions). For small $n$, there is a second order effect that penalizes designs with small block size.
\end{remark}

\section{Simulations}\label{sec:simulations}

\subsection{One Covariate}\label{subsec:simulation_p_1}

We first provide simulation evidence for our flagship result (Remark~\ref{rem:pm_optimal}) when the ordering of the entries of $\v$ is known prior to experimental allocation. We assume the typical generalized linear model with additive treatment effect where the probability that $y=1$ is log-odds-linear, i.e., $\beta_0 + \beta_1 x_i + \beta_T w_i$. As log odds is one-to-one with the average probability $\v$ of Definition~\ref{eq:def_of_v_vec}, the order of $v_i$'s is the same as the order of the $x_i$'s. Here, blocking into $B$ strata involves sorting $x$ and slicing by sample quantiles $1/B, \ldots (B-1)/B$ and PM's optimal pair structure is the indices order corresponding to the pairs $<x_{(1)}, x_{(2)}>, \ldots, <x_{(2n - 1)}, x_{(2n)}>$ where $x_{(i)}$ denotes the $i$th smallest $x$ value. 

As the CMHT is asymptotic and our results are asymptotic, we wish to understand performance across a range of sample sizes realistic in clinical trials and we consider all numbers of blocks of even size (which we require for the simulation in Section~\ref{subsec:simulation_p_gr_1}) with maximum of $B=n$ for PM and where $B$ divides $2n$. These constraints result in 74 cells over all values of $2n \in \braces{48, 96, 192, 288, 384}$, values specifically selected as they have many even factors. 

For each value of $2n$, the covariates are drawn once from an iid standard normal. We then set $\beta_0 = 0$ and $\beta_1 = 1$ and $\beta_T \in \braces{0, 0.7}$ resulting in 74 $\times$ 2 = 148 cells. The setting of $\beta_T = 0$ allows us to gauge if the CMHT is properly sized over all sample size / block size cells and the setting of $\beta_T = 0.7$ was selected in order to induce separation in power measurements over all sample size / block size cells. As the $x_i$'s are drawn for each value of $2n$, for each value of $\beta_T$ we compute the static potential probabilities of outcomes $p_{T_i}$ and $p_{C_i}$ for all subjects.

For all 148 cells, we simulate $N_Y := 1,000,000$ different assignments $\w$ drawn from their appropriate block design. Each of these $\w$'s selects an alternating spliced subset of the two potential probabilities of outcomes resulting in the $2n$ probabilities of $y_i = 1$. We then draw the response values of using these probabilities from independent heterogeneous Bernoullis. That is, for each of the $N_Y$ draws, we have the realized responses $\y$, the assignment $\w$ and the strata via the block structure. Thus, we can calculate the MH statistic (of Equation~\ref{eq:CMHT_statistic}) and then we can decide the null at $\alpha = 5\%$ (i.e, if $MH > 3.84$, the critical $\chi^2_1$ threshold, the null is rejected).

\begin{figure}[htp]
    \centering
    \includegraphics[width=\linewidth]{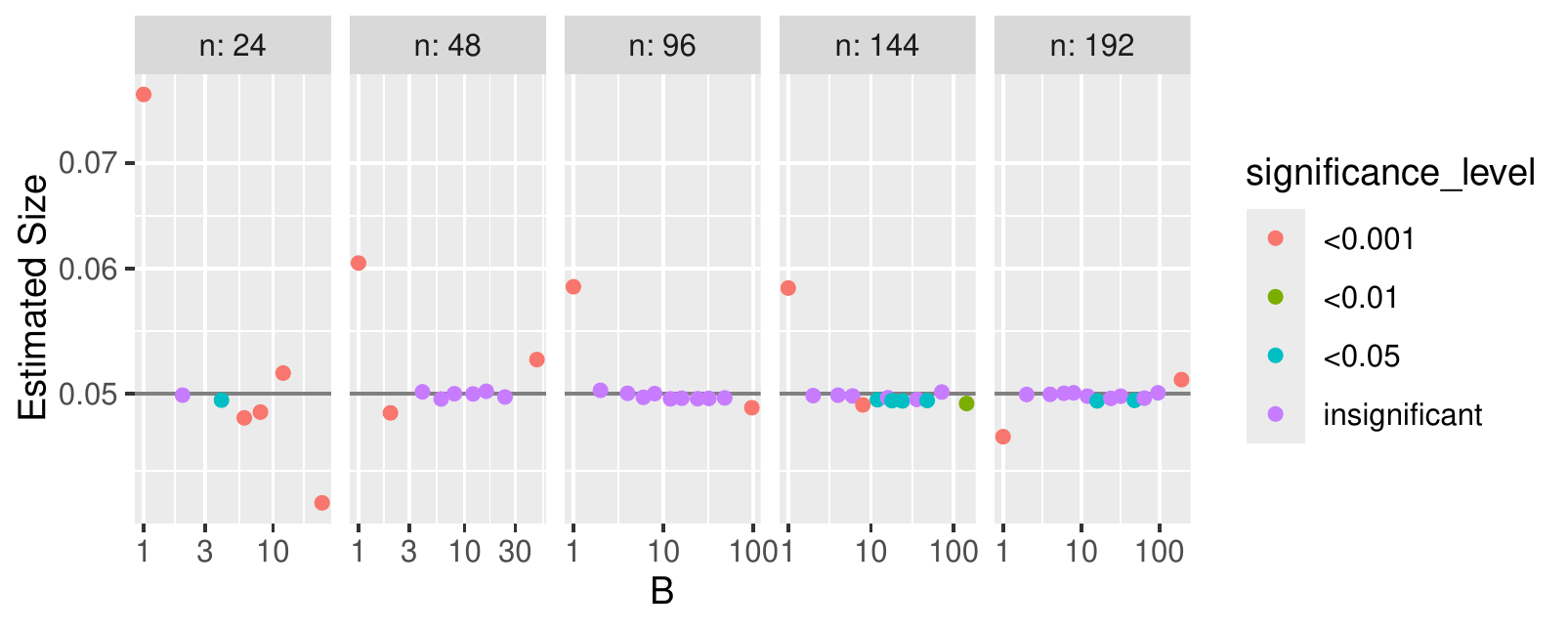}
    \caption{Size results with significance level for all 74 sample size cross block designs considered. Confidence intervals are smaller than the dots and thus unshown. \qu{Significance level} is displayed for a one-proportion two-sided z-test where $H_0: \alpha = 5\%$.}
    \label{fig:sizes_p_1}
\end{figure}

The results to gauge size of the CMHT (Bonferroni-corrected for all simulation cells described here plus those in Section~\ref{subsec:simulation_p_gr_1}) are found in Figure~\ref{fig:sizes_p_1}. For $2n=48$, most cells are inappropriately-sized, which is to be expected due to low sample size; for $2n=96$, fewer cells are inappropriately-sized; and for $2n \geq 192$, most cells are appropriately sized. Although we find statistically significant violations of size $\neq \alpha$, the estimated sizes themselves are approximately $\alpha$ with the only serious departure for $B=1$ (BCRD) across all sample sizes. However, BCRD is not what the CMHT was designed for (as it lacks strata).

\begin{figure}[htp]
    \centering
    \includegraphics[width=\linewidth]{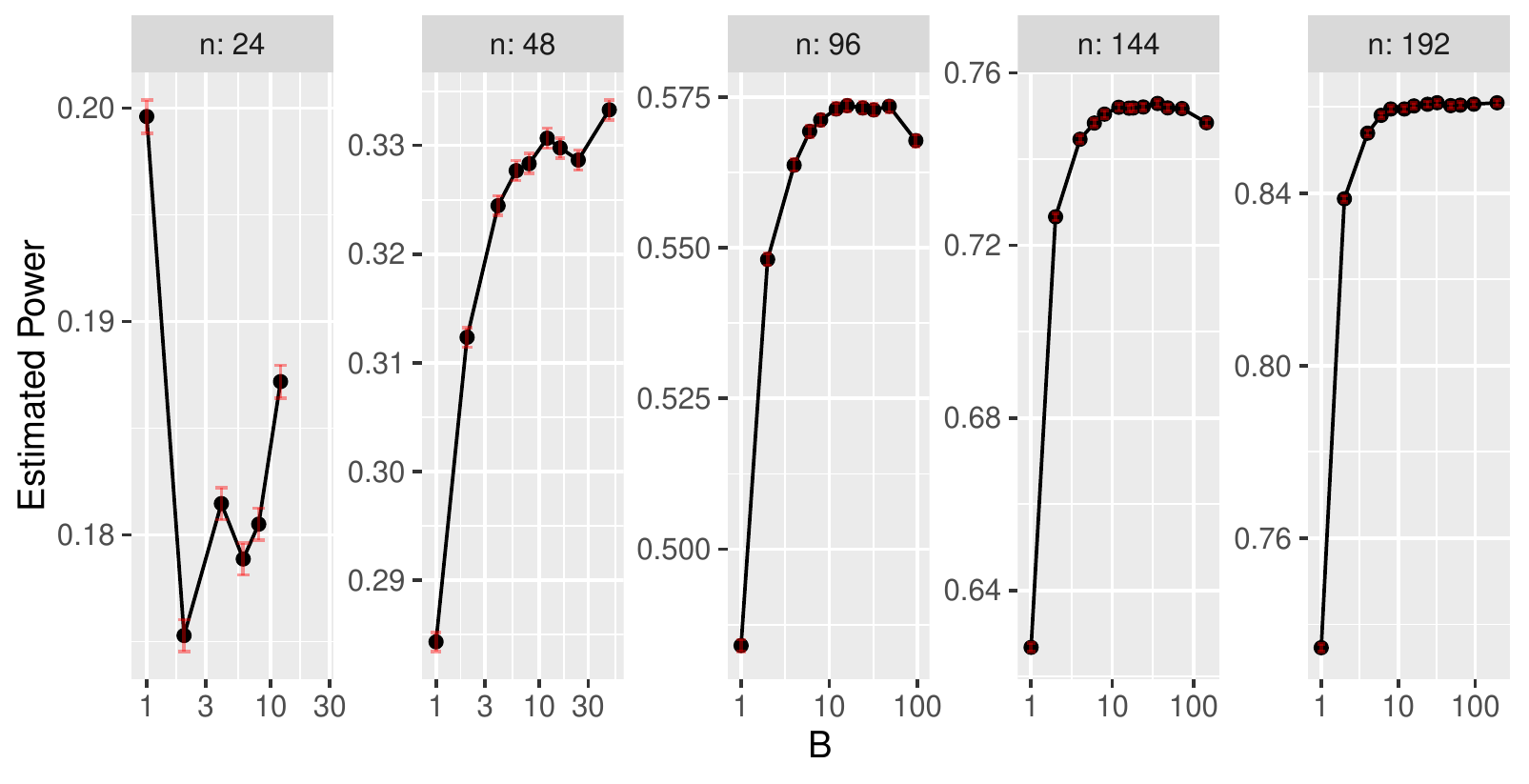}
    \caption{Power results for all 74 sample size cross block designs considered. The red error bars indicate 95\% confidence intervals constructed from the $N_Y$ independent simulations.}
    \label{fig:power_p_1}
\end{figure}

The power results are shown in Figure~\ref{fig:power_p_1}. In the smallest sample size $2n=48$, the power cannot be trusted due to the inappropriate sizing explained above. For $2n \in \braces{96, 192, 288}$, the first order term in $n$ mostly dominates but as $B$ increases, the second order term in $n$  becomes significant leading to a peak in $B$ and slight decrease up to PM. By the largest sample size $2n = 384$, the second order term vanishes and PM is observed to be optimal. This result is expected given the discussion about the second order term below Equation~\ref{eq:small_n_effect_eqs}.

\subsection{Multiple Covariates}\label{subsec:simulation_p_gr_1}

We now wish to explore the more realistic scenario of more than one covariate by simulating for $p \in \braces{2, 5}$. Here, the blocking structure becomes imperfect as it can no longer splice subjects along the ordered $\v$ (see Definition~\ref{eq:def_of_v_vec}). For $p=2$, we first sort along the first covariate, then we create $B/2$ strata at the sample quantiles. Then within each of these strata, we sort along the second covariate and divide into 2 strata. (This is the reason we required all $B$ to be even in the previous section to conveniently have the same cells across for all $p$). For $p=5$, we use the same procedure ignoring the third, fourth and fifth covariates. We view this as realistic as in practice; blocking on more than two covariates in small sample sizes is known to be difficult. And for convenience, we retain the same blocking procedures for the larger simulated sample sizes. 

The previous paragraph discusses how we block for $B < n$. Although this procedure theoretically allows for PM, in standard practice PM is handled categorically differently as we explain now. One first calculates the proportional between-subjects Mahalanobis distance of observed covariates \citep[Section 2.2]{Stuart2010} via $(\x_\ell - \x_m)^\top \hat{\Sigma}_X^{-1} (\x_\ell - \x_m)$ where $\hat{\Sigma}_X^{-1}$ is the $p \times p$ sample variance covariance matrix of the $2n$ subjects' covariate vectors for all the $\binom{2n}{n}$ pairs of subjects with indices $\ell, m$. This results in a symmetric distance matrix of size $2n \times 2n$. Then, one runs the optimal nonbipartite matching algorithm that executes in polynomial time \citep{Lu2011} which runs on the distance matrix returning the best-matched set of pairs of indices. This procedure provided a match structure that \textit{approximates} the optimal match structure corresponding to the subject indicies of $<v_{(1)}, v_{(2)}>, \ldots, <v_{(2n - 1)}, v_{(2n)}>$. We employ this procedure for $B=n$ and thus our simulation results in this subsection will be categorically different for PM when compared to $B<n$, but not in Section~\ref{subsec:simulation_p_1} (as this procedure would be the same as sorting on the one, sole covariate).

To draw the covariate values, we use the same procedure as in Section~\ref{subsec:simulation_p_1}, i.e., iid standard normal realizations. The linear model is now expanded for $p$ covariates and we set $\beta_j = 1$ for all covariates. The value of $\beta_0$ and the value of $\beta_T$ we keep the same as in Section~\ref{subsec:simulation_p_1}. We use the same sampling procedure for $N_Y = 1,000,000$ draws in each of the $74 \times 2 \times 2 = 296$ new cells.

\begin{figure}[htp]
    \centering
    \includegraphics[width=\linewidth]{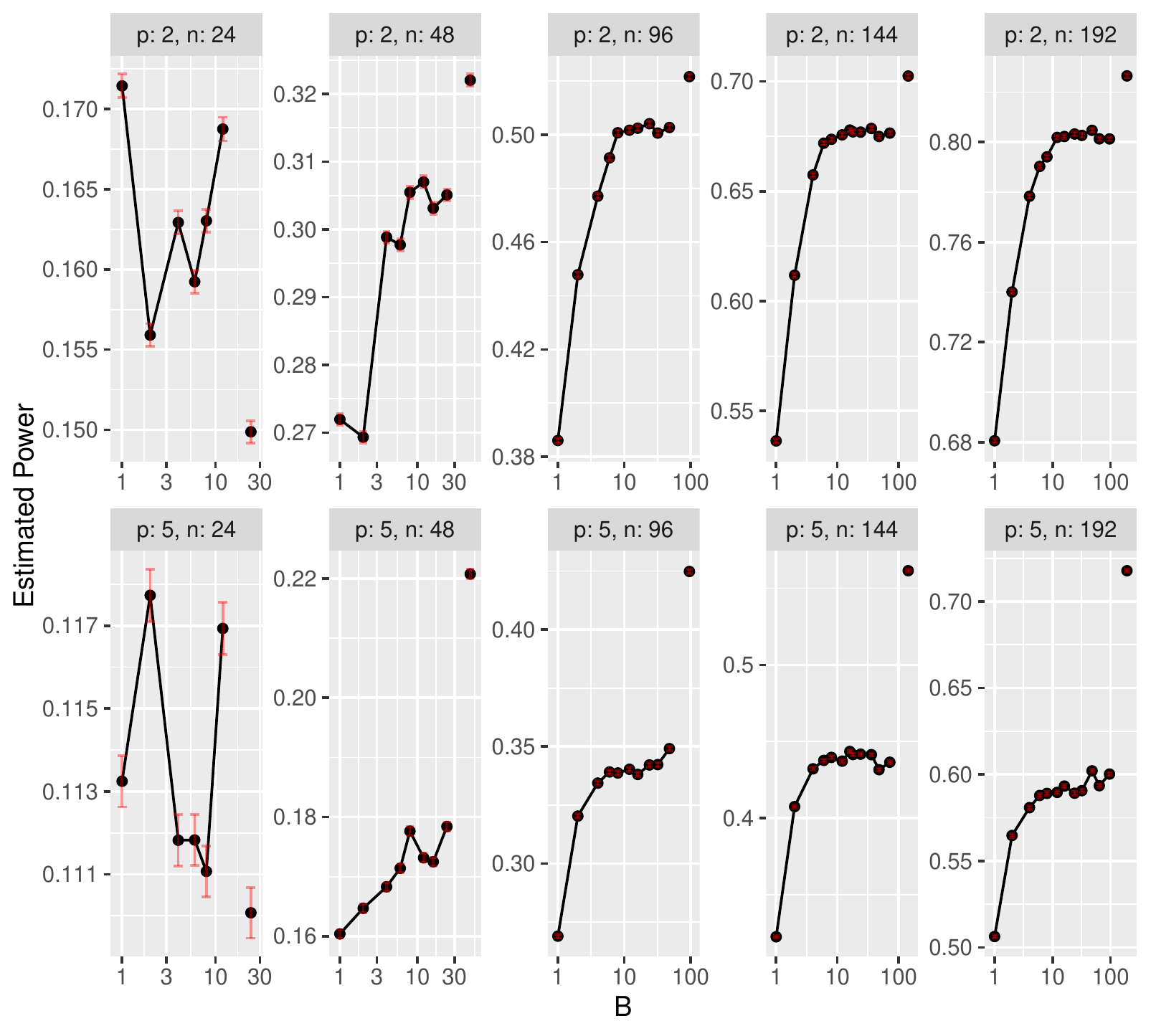}
    \caption{Power results for all 148 sample size cross block designs cross number of covariates considered. The error bars are the same as in Figure~\ref{fig:power_p_1}. The disconnected dot found at $B=n$ is the performance of PM.}
    \label{fig:power_p_2_5}
\end{figure}

The Bonferroni-corrected size results are found in Figure~\ref{fig:sizes_p_gr_1} in Section B of the Supporting Information and the results are similar to $p=1$ with a bit more violations when $p=5$. The power results are shown in Figure~\ref{fig:power_p_2_5} where the PM results are shown as line-disconnected dots (as the PM results are categorically different than the $B<n$ designs). We observe largely the same pattern as in Figure~\ref{fig:power_p_1} for the $B<n$ blocking designs. But PM now clearly has the highest power as it is able to most-closely create the optimal match structure. This is akin with our flagship result (Remark~\ref{rem:pm_optimal}) as PM maximizes the asymptotic power even when the blocking is not perfect.

\section{Discussion}\label{sec:discussion}

We examined power in the two-arm blocking design with a binary response. The $B$ blocks can be thought of as different strata and our data is then arranged as a $2 \times 2 \times B$ array of counts. We consider the CMHT with the null hypothesis of equal odds ratio of positive outcome across both arms. We then analyze its power function over local alternatives with designs of differing number of blocks. We find that under perfect blocking (i.e., knowledge of the order of potential outcomes' probabilities) the asymptotic power is maximized for all blocking design with $B\to \infty$. For non-perfect blocking we show that PM maximizes the asymptotic power under mild assumptions. 

In small sample sizes, there is a second-order drag term that becomes significant and thus the optimal design is a large number of blocks, but not as many as PM. Our simulations with realistic sample sizes show the CMHT under block designs (with more than one block) is mostly appropriately-sized which means the asymptotics for the CMHT engage relatively quickly. Simulated power estimates support our theoretical results for one covariate. For more than one covariate, we cannot optimally block (nor pairwise match) based on the order of the known probabilities of the potential outcomes' probabilities. Here, PM is the clear best performer because it maximizes the asymptotic power and it
achieves the closest approximation of matches on the potential outcomes' probabilities. For this practical point, it is our design recommendation.

There are many further avenues to explore. We only examined the power of the CMHT; we did not explore mean squared error of the estimator nor confidence interval coverage. Our intuition is that the resultant optimal error and optimal coverage design will likely be the same. However, it is theoretically difficult to have tractability on estimators for the odds ratio and log odds ratio.  Although one typically uses logistic regression for estimation and testing in our experimental setting, it assumes a linear relationship which is not assumed here. The logistic regression estimator is also difficult to analyze as the Fisher Information depends on the unknown linear coefficients. Also, we analyzed the non-sequential (or \qu{offline}) setting that was originally studied by \citet{Fisher1925} when assigning treatments to agricultural plots. The more common \emph{sequential}  (or \qu{online}) experimental setting where subjects arrive one-by-one and must be assigned a $w_i$ on the spot (hence all $\x_i$'s are \textit{not} seen beforehand) is an interesting setting to explore.

\section*{Acknowledgments}

This research was supported by Grant No 2018112 from the United States-Israel Binational Science Foundation (BSF). We wish to thank Yosi Rinott for the help with the proof of our flagship result.

\pagebreak


\pagebreak
\clearpage
\appendix

\begin{center}
    \Large{Supporting Information for\\ \qu{\ourtitle}}\\~\\
    \footnotesize{by David Azriel, Adam Kapelner \& Abba M. Krieger}
\end{center}

\refstepcounter{section}\addcontentsline{toc}{section}{\protect\numberline{\thesection}A}

\subsection{CMHT Statistic Calculation}\label{sec:calc}
We wish to show that

\[
MH := \frac{
\squared{\,\displaystyle\sum_{b=1}^B \left( n_{T,1_b} -\frac{n_{1_b}}{2} \right)}
}{
\displaystyle\oneover{4(n_B-1)}\sum_{b=1}^B n_{1_b} n_{0_b}
} = \frac{(\W^\top \Y)^2  }{ \Y^\top \bSigmawB \Y }.
\]

Let $I_b$ be the indices of subjects in the $b$-th block. 
Then,

\[
n_{T,1_b}=\sum_{i \in I_b} I(W_{B,i}=1) Y_i,~n_{1_b}=\sum_{i \in I_b} I(W_{B,i}=1) Y_i.
\]

\noindent Therefore,

\[
 n_{T,1_b} -\frac{n_{1_b}}{2}=\sum_{i \in I_b} [ I(W_{B,i}=1) Y_i-Y_i/2].
\]

\noindent Notice that if $W_{B,i}=1$, then $I(W_{B,i}=1) Y_i-Y_i/2 = Y_i/2$ and if $W_{B,i}=-1$, then $I(W_{B,i}=1) Y_i-Y_i/2 = -Y_i/2$. Thus, $I(W_{B,i}=1) Y_i-Y_i/2 =W_{B,i} Y_i/2$. Hence,

\[
 n_{T,1_b} -\frac{n_{1_b}}{2}=  \sum_{i \in I_k} W_{B,i} Y_i/2.
\]

\noindent Therefore, the numerator of the MH test statistics in Equation \ref{eq:CMHT_statistic} is $\frac{1}{4} (\w^\top \y)^2$. Consider now the denominator of MH in Equation \ref{eq:CMHT_statistic}.
We have that

\[
 n_{1_b} n_{0_b} = \left( \sum_{i \in I_b} y_i \right) \left( \sum_{i \in I_b}(1- y_i) \right)= n_B\sum_{i \in I_b} Y_i - \left( \sum_{i \in I_b} Y_i\right)^2, 
\]

\noindent where in the last equality we used the relation $Y_i=Y_i^2$, which is true because $Y_i \in \{0,1\}$. Furthermore,

\[
 n_B\sum_{i \in I_b} y_i - \left( \sum_{i \in I_b} y_i\right)^2= n_B  \sum_{i \in I_b} \left(  Y_i- \bar{Y}_b \right)^2,
\]

\noindent where $\bar{Y}_b = \frac{1}{n_B} \sum_{i \in I_b} Y_i$. It follows that denominator of MH in Equation \ref{eq:CMHT_statistic} is

\[
\oneover{4(n_B-1)}\sum_{b=1}^B n_{1_b} n_{0_b} = \frac{n_B}{4(n_B-1)}\sum_{b=1}^B \sum_{i \in I_b} \left(  Y_i- \bar{Y}_b \right)^2.
\]

\noindent In our previous paper's Supporting Information \citep[Appendix A.2, Equation 18]{Kapelner2023} we showed that $\frac{n_B}{n_B-1}\sum_{b=1}^B \sum_{i \in I_b} \left(  Y_i- \bar{Y}_b \right)^2=\Y^\top \bSigmawB \Y$. Hence,

\[
MH=\frac{  (\W_B^\top \Y)^2  }{ \Y^\top \bSigmawB \Y }.
\]

\subsection{Proof of Main Result}\label{sec:proof}


By Lemma \ref{lem:SB} $\frac{\frac{1}{2n} \W_B^\top\Y - \bar{\tau}_n }{\sqrt{\eta_n/2n}}$ converges in distribution to $N(0,1)$ under {\em small blocks asymptotics} and Lemma \ref{lem:LB} implies that the same convergence holds under {\em large blocks asymptotics}. Notice that we assumed that $\tilde{v}_0$ is bounded from below. By Lemma \ref{lem:quad},   $\frac{1}{2n} \Y^\top \bSigmawB \Y - \expe{\frac{1}{2n} \Y^\top \bSigmawB \Y}$ converges in $L_2$, and hence in probability, to zero. Also, by Lemma \ref{lem:quad},
\begin{equation*}
\expe{ \frac{1}{2n}  \Y^\top \bSigmawB \Y}
=\frac{1}{2n} \left[  \v^\top \bSigmawB \v + \sum_{i=1}^{2n} v_i(1-v_i) + \frac{1}{n_B-1} \left( \normsq{\btau} - {\boldsymbol \tau}^\top \bSigmawB {\boldsymbol \tau}   \right) \right].
\end{equation*}
The maximal eigenvalue of $\bSigmawB$ is bounded by 2 (which is the maximal eigenvalue for pairwise matching, i.e., when $B=n$). Hence, ${\boldsymbol \tau}^\top \bSigmawB {\boldsymbol \tau} \le \normsq{\btau}$. We assumed that $\lim_{n \rightarrow \infty} \frac{1}{2n}\normsq{\btau}=0$ and therefore, 
\[
\lim_{n \rightarrow \infty} \frac{1}{2n}\left( \normsq{\btau} - {\boldsymbol \tau}^\top \bSigmawB {\boldsymbol \tau}   \right)=0.
\]
It follows that $\expe{ \frac{1}{2n}  \Y^\top \bSigmawB \Y} - \eta_n \to 0$. Hence,
$\frac{1}{2n}  \Y^\top \bSigmawB \Y$ converges in probability to $\eta_\infty$. 

We have that

\bneqn \label{eq:tMH}
\widetilde{MH}= \frac{\frac{1}{2n} \W_B^\top \Y }{ \sqrt{\frac{1}{2n} \Y^\top \bSigmawB \Y}/\sqrt{2n} } = \frac{\frac{1}{2n} \W_B^\top \Y - \bar{\tau}_n }{ \sqrt{\frac{1}{2n} \Y^\top \bSigmawB \Y}/\sqrt{2n} }+\frac{\bar{\tau}_n \sqrt{2n}}{ \sqrt{\frac{1}{2n} \Y^\top \bSigmawB \Y} }. 
\eneqn

\noindent We showed that $\frac{\frac{1}{2n} \W_B^\top\Y - \bar{\tau}_n }{\sqrt{\eta_n/2n}}$ converges in distribution to $N(0,1)$ and that $\frac{1}{2n}  \Y^\top \bSigmawB \Y$ converges in probability to $\eta_\infty$. By Slutsky's theorem, $\frac{\frac{1}{2n} \W_B^\top \Y - \bar{\tau}_n }{ \sqrt{\frac{1}{2n} \Y^\top \bSigmawB \Y}/\sqrt{2n} }$ converges in distribution to $N(0,1)$. Since 
$\frac{1}{2n}  \Y^\top \bSigmawB \Y$ converges in probability to $\eta_\infty$, the left term in \eqref{eq:tMH} converges in probability to $c/\eta_\infty$. The result of the theorem follows.

\begin{lemma}\label{lem:SB}
Consider the model of Section \ref{sec:setup} and assume that $\eta_n \to \eta_\infty>0$, we have for a constant $C$ that
\[
\sup_t \left| \prob{\frac{\frac{1}{2n} \W_B^\top\Y - \bar{\tau}_n }{\sqrt{\eta_n/2n}} \le t}-\Phi(t)  \right| \le C \sqrt{n}/B. 
\] 
\end{lemma}

\noindent{\bf Proof of Lemma \ref{lem:SB}.} We can write 
\[
\frac{1}{2n} \W_B^\top \Y - \bar{\tau}_n = \frac{1}{B} \sum_{b=1}^B \frac{1}{n_B} \sum_{i \in I_b} (W_i Y_i - \tau_i ).
\]
For each $b$, $\frac{1}{n_B} \sum_{i \in I_b} \parens{W_i Y_i - \tau_i  }$ has mean zero and variance $ \frac{\eta_b}{n_B}$ where
\[
\eta_b:=  \frac{1}{n_B-1} \sum_{i \in I_b}(v_i- \bar{v}_b)^2+ \frac{1}{2 n_B} \sum_{i \in I_b} \left\{ p_{T,i}(1-p_{T,i})+p_{C,i}(1-p_{C,i}) \right\}. 
\]
The third moment of $\frac{1}{n_B} \sum_{i \in I_b} \parens{W_i Y_i - \tau_i  }$ can be bounded as follows
\bneqn\label{eq:bound_3_moment}
\expe{\left|\frac{1}{n_B} \sum_{i \in I_b} \parens{W_i Y_i - \tau_i  }\right|^3}\le 2 \expe{\frac{1}{n_B} \sum_{i \in I_b} \parens{W_i Y_i - \tau_i  }}^2=2\var{\frac{1}{n_B} \sum_{i \in I_b} \parens{W_i Y_i - \tau_i  }}=\frac{2 \eta_b}{n_B},
\eneqn
where the inequality is true because $|W_i Y_i - \tau_i|\le 2$ for all $i$.

We now invoke the Berry–Esseen inequality using the notation of \citet[Theorem 3.6]{Chen2010}. Recall the definition of $\eta_n$ given in Equation~\ref{eq:def_of_eta_vec}. It follows that
 $\eta_n=\frac{n_B}{2n} \sum_{b} \eta_b=\frac{1}{B}\sum_{b} \eta_b$. Let $\xi_b:= \frac{\frac{1}{n_B} \sum_{i \in I_b} \parens{W_i Y_i - \tau_i  }}{\sqrt{\eta_n B/n_B}}$. Then $\xi_1,\ldots,\xi_B$ are independent with mean zero and
\[
\sum_{b=1}^B \var{\xi_b}=\frac{\frac{1}{n_B}\sum_{b=1}^B \eta_B}{ B \eta_n/n_B}=
\frac{\frac{1}{B}\sum_{b=1}^B \eta_B}{  \eta_n}=1.
\]
Furthermore,
\[
\sum_{b=1}^B \xi_B=\frac{\frac{1}{\sqrt{B n_B }} \sum_{b=1}^B \sum_{i \in I_b} \parens{W_i Y_i - \tau_i  } }{\sqrt{\eta_n}} 
= \frac{\frac{1}{2n} \sum_{b=1}^B \sum_{i \in I_b} \parens{W_i Y_i - \tau_i  } }{\sqrt{\eta_n/2n}}=
\frac{\frac{1}{2n} \W_B^\top\Y - \bar{\tau}_n }{\sqrt{\eta_n/2n}}.
\]
By Theorem 3.6 of \cite{Chen2010},
\begin{equation}\label{eq:xi_b}
\sup_t \left| \prob{\sum_{b=1}^B \xi_B \le t}-\Phi(t)  \right| \le C \sum_{b=1}^B \expe{ |\xi_b|^3}.     
\end{equation}
We assumed that $\eta_n \to \eta_\infty>0$ and therefore, $\eta_n$ is bounded from below; also, by Equation~\ref{eq:bound_3_moment}, the third moment of $\frac{1}{n_B} \sum_{i \in I_b} \parens{W_i Y_i - \tau_i  }$ can be bounded by $2\eta_B/n_B \le C/n_B$. It follows that $\expe{ |\xi_b|^3} \le C \frac{1/n_B}{\parens{B/n_B}^{3/2}}$.
Therefore,
\[
\sum_{b=1}^B \expe{ |\xi_b|^3} \le  
C\frac{B/n_B}{\parens{B/n_B}^{3/2}}= C \sqrt{n_B/B}=\sqrt{2}C \sqrt{n}/B.
\]
Hence, Equation~\ref{eq:xi_b} implies the reult.
\qed

\begin{lemma}\label{lem:LB}
Consider the model of Section \ref{sec:setup}, we have for a constant $C$ that
\[
\sup_t \left| \prob{\frac{\frac{1}{2n} \W_B^\top\Y - \bar{\tau}_n }{\sqrt{\eta_n/2n}} \le t}-\Phi(t)  \right| \le \frac{C}{ \tilde{v}_0}\parens{ \frac{1} {\sqrt{n}} +  \frac{\sigma_{max}}{\sigma_{min}} \frac{B}{\sqrt{n}}  }, 
\] 
where $\tilde{v}_0=\frac{1}{4n} \sum_{i=1}^{2n} \parens{p_{T,i}(1-p_{T,i})+p_{C,i}(1-p_{C,i})}$.
\end{lemma}
\noindent{\bf Proof of Lemma \ref{lem:LB}.}
We first condition on $\W_B=\w_0$ for certain $\w_0$. Then, $w_{0,i} Y_i - \tau_i$ are independent with mean $w_{0,i} v_i$ and variance $\tilde{v}_i^2 := \frac{1}{2} \parens{p_{T,i}(1-p_{T,i})+p_{C,i}(1-p_{C,i})}$. Let $\tilde{v}_0^2 :=\frac{1}{2n}  \sum_{i=1}^{2n} \tilde{v}_i^2$. Similar to the argument in Lemma \ref{lem:SB}, since $w_{0,i} Y_i - \tau_i$ is bounded, $\expe{\abss{w_{0,i} Y_i - \tau_i}^3}\le 2 \tilde{v}_i^2$. 
Hence, Theorem 3.6 of \citet{Chen2010} implies that
\[
 \left| \cprob{ \frac{\frac{1}{2n} \W_B^\top \Y - \bar{\tau}_n}{\sqrt{\eta_n/2n}} \le t}{\w=\w_0} -\Phi\left(\frac{t \sqrt{\eta_n/2n}  -\frac{1}{2n} \w_0^\top \v}{ \tilde{v}_0\sqrt{\frac{1}{2n} }} \right)  \right| \le \frac{C}{\sqrt{2n} \tilde{v}_0 } . 
\]
Since it is the same bound for each $\w_0$, it follows that
\begin{equation}\label{eq:expr.w}
 \left| \prob{ \frac{\frac{1}{2n} \W_B^\top \Y - \bar{\tau}_n}{\sqrt{\eta_n/2n}} \le t} -\expe{\Phi\left(\frac{t \sqrt{\eta_n/2n} -\frac{1}{2n} \W_B^\top \v}{ \tilde{v}_0\sqrt{\frac{1}{2n} }} \right)}  \right| \le  \frac{C}{\sqrt{2n} \tilde{v}_0 }  .    
\end{equation}

In order to establish asymptotic normality we need to show that $\frac{1}{2n} \W_B^\top \v$ is asymptotically normal. We have that $\W_B^\top \v = \sum_{b=1}^B \W_b^\top \v_b$, where $\W_b$ and $\v_b$ are sub-vectors corresponding to the $b$-th block. Notice that the blocks are independent. 

We will now invoke Theorem 4.8 of \citet{Chen2010} for each term $\W_b^\top \v_b$. In the notation of the latter theorem we consider a matrix $\A$ of dimension $n_B \times n_B$, whose $i$-th row is equal to $(\onevec_{n_b/2}^\top v_{b,i},-\onevec_{n_b/2}^\top v_{b,i} )$.
Thus, $\W_b^\top \v_b$ has the same distribution of $\sum_{i=1}^{n_b} A_{i,\pi(i)}$, where $\pi$ is a permutation with uniform distribution over the permutation group.

We have that $\expe{ \W_b^\top \v_b}=0$ and $\var{\W_b^\top \v_b}=\sigma^2_b:=\frac{n_b}{n_b-1}\sum_{i\in I_b}(v_{i,b} - \bar{v}_b)^2$. Since $|W_{b,i}v_i| \le 1$ for all $i \in I_b$ it follows that
$\expe{ \left|\W_b^\top \v_b\right|^3}\le n_b \sigma_b^2$.
Thus, we have by  Theorem 4.8 of \citet{Chen2010} that

\beqn
\abss{\expe{h\parens{\frac{\W_b^\top \v_b}{\sqrt{\var{ \W_b^\top \v_b}}}}} - \expe{h(Z)}} \le \frac{C}{\sigma_b}.
\eeqn

\noindent for $Z\sim N(0,1)$ and $h$ a Lipschitz function. When $h$ is differentiable, then $h'(z)\le 1$ for all $z$. 
Since the blocks are independent and $\W_B^\top \v = \sum_{b=1}^B \W_b^\top \v_b$, we have by Lemma \ref{lem:U_b} below that
\begin{equation}\label{eq:expr.h}
\abss{\expe{h\parens{\frac{\W_B^\top \v}{\sqrt{\var{ \W_B^\top \v}}}}} - \expe{h(Z)}} \le  \frac{C }{\sqrt{B}} \sum_{b=1}^B \frac{1}{\sigma_b}.    
\end{equation}

\noindent We have that $\var{ \W_B^\top \v}=\v^\top \bSigmaw \v=\sum_b \sigma^2_b$. Going back to Equation~\ref{eq:expr.w}, define

\[
h_0\left( z\right) := \Phi\left(\frac{t \sqrt{\eta_n/2n}-z \frac{\sqrt{\sum_b \sigma^2_b }}{2n} }{ \tilde{v}_0\sqrt{\frac{1}{2n} }} \right) = \Phi\left(\frac{\sqrt{2n \eta_n}t -z \sqrt{\sum_b \sigma^2_b}}{\tilde{v}_0\sqrt{2n} } \right) 
\]

\noindent for $z=\W_B^\top \v/ \sqrt{\sum_b \sigma^2_b}$. Note that its derivative is bounded by

\[
|h_0'\left( z\right)|= \varphi\left(\frac{\sqrt{2n \eta_n}t -z \sqrt{\sum_b \sigma^2_b}}{\tilde{v}_0\sqrt{2n} }\right)  \frac{\sqrt{\sum_b \sigma^2_b}}{\tilde{v}_0\sqrt{2n} } \le C \frac{\sqrt{\sum_b \sigma^2_b}}{ \tilde{v}_0 \sqrt{2n} }. 
\]

\noindent Therefore, Equation~\ref{eq:expr.h} implies that

\begin{multline}\label{eq:boundd}
\left| \expe{\Phi\left(\frac{t \sqrt{\eta_n/2n}-\frac{1}{2n} \W_B^\top \v}{ \tilde{v}_0\sqrt{\frac{1}{2n} }} \right)}- \expe{\Phi\left(\frac{\sqrt{2n \eta_n}t-Z \sqrt{\sum_b \sigma^2_b}}{\tilde{v}_0\sqrt{2n} } \right) }\right|\\
\le   \frac{C }{\sqrt{B}} \sum_{b=1}^B \frac{1}{\sigma_b} \times \frac{\sqrt{\sum_b \sigma^2_b}}{ \tilde{v}_0 \sqrt{2n} }.     
\end{multline}

\noindent Now,
\[
\expe{\Phi\left(\frac{\sqrt{2n \eta_n}t -Z \sqrt{\sum_b \sigma^2_b}}{\tilde{v}_0\sqrt{2n} } \right) } = \prob{ \tilde{Z} \le \frac{\sqrt{2n \eta_n}t -Z \sqrt{\sum_b \sigma^2_b}}{\tilde{v}_0\sqrt{2n} }   },     
\]
where $\tilde{Z}\sim N(0,1)$ is independent of $Z$. Hence,
\[
\prob{ \tilde{Z} \le \frac{\sqrt{2n \eta_n}t -Z \sqrt{\sum_b \sigma^2_b}}{\tilde{v}_0\sqrt{2n} }   }= \prob{\frac{ \tilde{Z}  \tilde{v}_0\sqrt{2n} + Z \sqrt{\sum_b \sigma^2_b} }{\sqrt{ 2n \eta_n}} \le t  } =\Phi(t ), 
\]
because
\[
2n \eta_n=\v^\top \bSigmawB \v + \frac{\p_T^\top (\onevec - \p_T)+\p_C^\top (\onevec - \p_C)}{2} = \sum_b \sigma^2_b+ 2n \tilde{v}_0^2.
\]
Now, Equations~\ref{eq:boundd} and \ref{eq:expr.w} imply that
\[
\sup_t \left| \prob{ \frac{\frac{1}{2n} \W_B^\top \Y - \bar{\tau}_n}{\sqrt{\eta_n/2n}} \le t} -\Phi(t)  \right| \le  \frac{C}{  \tilde{v}_0\sqrt{2n}} \parens{ 1 +  \frac{{\sqrt{\sum_b \sigma^2_b}}}{\sqrt{B}} \sum_{b=1}^B \frac{1}{\sigma_b}  } \le  \frac{C}{  \tilde{v}_0} \parens{ \frac{1} { \sqrt{2n}} +  \frac{\sigma_{max}}{\sigma_{min}} \frac{B}{\sqrt{2n}}  }.
\]
\qed

\begin{lemma}\label{lem:U_b}
Suppose that 
\[
\abss{\expe{h(U_b)}-\expe{h(Z)}} \le C_b \text{ for }b=1\ldots,B,
\]
where $U_1,\ldots,U_B$ are independent with mean zero and variance 1 and $Z\sim N(0,1)$, and the bound holds for all $h$ differentiable with $|h'(z)|\le 1$ for all $z\in \mathbb{R}$. Then,
\[
\left|\expe{h\left(\frac{1}{\sqrt{B}}\sum_{b=1}^B U_b\right)}-\expe{h(Z)} \right| \le \frac{\sum_{b=1}^B C_b}{\sqrt{B}}
\]
 for all $h$ as above.
 \end{lemma}
 
 \noindent{\bf Proof of Lemma \ref{lem:U_b}.} We first condition on $U_2=u_2,\ldots,U_B=u_B$. For each function $h$, define $\tilde{h}(u):=h\left(\frac{1}{\sqrt{B}}\left(u+\sum_{b=2}^B u_b\right)\right)$. We have that $|\tilde{h}'(u)| \le \frac{1}{\sqrt{B}}$ for all $u$ and $u_2,\ldots,u_B$. Therefore, by the above bound for $U_1$,
 \[
 \left|\cexpe{\tilde{h}(U_1)}{U_2=u_2,\ldots,U_B=u_B}-\cexpe{\tilde{h}(Z)}{U_2=u_2,\ldots,U_B=u_B}\right| \le \frac{C_1}{\sqrt{B}}.
 \] 
 Since the same bound holds for all $u_2,\ldots,u_n$, we have by the law of total expectation that
 \[
 \left|\expe{h\left(\frac{1}{\sqrt{B}}\sum_{b=1}^B U_b\right)}- \expe{ h\left(\frac{1}{\sqrt{B}}\left(Z_1+\sum_{b=2}^B U_b\right)\right)}\right| \le \frac{C_1}{\sqrt{B}},
 \] 
 for $Z_1 \sim N(0,1)$ that is independent from $U_1,\ldots,U_n$.
Continuing in a similar fashion for $U_2,\ldots,U_n$ we obtain
 \[
\left|\expe{h\left(\frac{1}{\sqrt{B}}\sum_{b=1}^B U_b\right)}- \expe{ h\left(\frac{1}{\sqrt{B}}\sum_{b=1}^B Z_b\right)} \right| \le \frac{\sum_{b=1}^B C_b}{\sqrt{B}},
\] 
where $Z_1, \ldots,Z_b$ are iid N(0,1). Since 
$\frac{1}{\sqrt{B}}\sum_{b=1}^B Z_b \sim N(0,1)$ the lemma follows. \qed

\begin{lemma}\label{lem:quad}
Consider the model of Section \ref{sec:setup}, then
\[
\expe{ \frac{1}{2n}  \Y^\top \bSigmawB \Y}
=\frac{1}{2n} \left[  \v^\top \bSigmawB \v + \sum_{i=1}^{2n} v_i(1-v_i) + \frac{1}{n_B-1} \left( \normsq{\btau} - {\boldsymbol \tau}^\top \bSigmawB {\boldsymbol \tau}   \right) \right].
\]
and we have for a constant $C$ that
\[
\var{ \frac{1}{2n} \Y^\top \bSigmawB \Y} \le C/n,
\]
\end{lemma}

\noindent{\bf Proof of Lemma \ref{lem:quad}.}
We have that
\[
\expe{ \frac{1}{2n} \Y^\top \bSigmawB \Y}=\frac{1}{2n} \left( \sum_{i_1 \ne i_2} \expe{Y_{i_1} Y_{i_2}} (\bSigmawB)_{i_1,i_2}+\sum_i \expe{Y_{i}^2}  \right)
\]
Suppose that $i_1,i_2$ belong to the same block (otherwise $(\bSigmawB)_{i_1,i_2}=0$). Then 
\[
\expe{Y_{i_1} Y_{i_2}}= \cexpe{Y_{i_1} Y_{i_2}}{\W_B}.
\]
Given $\W_B$, $Y_1$ and $Y_2$ are independent and $\cexpe{Y_i}{W_{B,i}}=v_i + W_{B,i} \tau_i$. Therefore,
\[
\cexpe{Y_{i_1} Y_{i_2}}{\W_B}= v_{i_1} v_{i_2} + \expe{W_{B,i_1} W_{B,i_2}}\tau_{i_1} \tau_{i_2}= (\bSigmawB)_{i_1,i_2} {\tau_{i_1} \tau_{i_2}}.
\]
Recall that when $i_1\ne i_2$ then $(\bSigmawB)_{i_1,i_2}=-\frac{1}{n_B-1}$ if $i_1,i_2$ belong to the same block and 0 otherwise (Lemma \ref{lem:moments}). Also,
\[
\expe{Y_i^2}= \expe{Y_i}= \expe{\cexpe{Y_i}{W_{B,i}}}=v_i.
\]
Hence,
\begin{multline*}
\expe{ \frac{1}{2n} \Y^\top \bSigmawB \Y}
=\frac{1}{2n} \left[ \sum_{i_1 \ne i_2} v_{i_1} v_{i_2} (\bSigmawB)_{i_1,i_2}+\sum_i v_i  -\frac{1}{n_B-1} \left( \sum_{i_1 \ne i_2} \tau_{i_1} \tau_{i_2} (\bSigmawB)_{i_1,i_2})\right) \right]\\
=\frac{1}{2n} \left[  \v^\top \bSigmawB \v + \sum_{i=1}^{2n} v_i(1-v_i) + \frac{1}{n_B-1} \left(  \normsq{\btau} - {\boldsymbol \tau}^\top \bSigmawB {\boldsymbol \tau}  \right) \right]. 
\end{multline*}
This shows the expectation part of the lemma. 

We next show that $\var{ \frac{1}{2n} \Y^\top \bSigmawB \Y} \le C/n$. 
We have that $\frac{1}{2n} \Y^\top \bSigmawB \Y $ is an average of of $B=2n/n_B$ independent blocks, where each block has bounded variance (since $Y_i$'s are bounded). Therefore, $\var{ \frac{1}{2n} \Y^\top \bSigmawB \Y} \le C/n$  when $n_B$ is bounded.
 
Consider now the case that $n_B \to \infty$. We have that 
\[
\var{ \frac{1}{2n} \Y^\top \bSigmawB \Y} = \expe{\cvar{ \frac{1}{2n} \Y^\top \bSigmawB \Y}{\w_B}} + \var{\cexpe{ \frac{1}{2n} \Y^\top \bSigmawB \Y}{\W_B}}.
\]
Given $\W_B$, the $Y_i$'s are independent. In our previous work \citep{Azriel2024} we have calculated $\var{\Z^\top \bSigmawB \Z}$ where $\Z$ has independent entries but not identically distributed (with mean zero). Based on these computations (see Eq. (32) in \citet{Azriel2024}) it can be shown that $\var{ \frac{1}{n} \Y^\top \bSigmawB \Y| \w} \le C/n$ for a constant $C$ when the moments of the $Y_i$'s and  $\frac{1}{n} \tr{\bSigmawB^2}$ are bounded, which is the case in our setting since $Y_i \in \{0,1\}$.

Consider now $ \var{\cexpe{ \frac{1}{2n} \Y^\top \bSigmawB \Y}{\W_B}}$. We have that
\begin{multline}\label{eq:Var_E}
 \var{\cexpe{ \frac{1}{2n} \Y^\top \bSigmawB \Y}{\W_B}}\\
 =\frac{1}{4n^2} \sum_{i_1,i_2,i_3,i_4}
 \Big\{ \cov{ (v_{i_1}+W_{B,i_1} \tau_{i_1})
(v_{i_2}+W_{B,i_2} \tau_{i_2})}{(v_{i_3}+W_{B,i_3} \tau_{i_3}) (v_{i_4}+W_{B,i_4} \tau_{i_4})}\\  (\bSigmawB)_{i_1,i_2} (\bSigmawB)_{i_3,i_4}\Big\}.
\end{multline}
If $i_1,i_2$ are not in the same block, then $(\bSigmawB)_{i_1,i_2}=0$; likewise $i_3,i_4$. If $i_1,i_2$ are not in the same block as $i_3,i_4$, then the covariance term in Equation~\ref{eq:Var_E} is zero. Therefore, 
it is enough to consider the case that $i_1,i_2,i_3,i_4$ all belong to the same block. We consider different cases:
\begin{itemize}
\item Suppose that $i_1=i_2=i_3=i_4$. There are order of $n$ such cases and therefore, 
 the sum of the covariances of this case is bounded by $C/n$.
\item Suppose that $i_1=i_2=i_3$ (or symmetrically $i_1=i_3=i_4$, $i_1=i_2=i_4$, $i_2=i_3=i_4$).  There are order of $\frac{n}{n_B} n_B^2= n n_B$ such cases. In this case $(\bSigmawB)_{i_1,i_1}=1$ and $(\bSigmawB)_{i_3,i_4}= -\frac{1}{(n_B-1)}$. Therefore, 
the sum of the covariances of this case is bounded by a contant times $n n_B \frac{1}{n^2} \frac{1}{(n_B-1)} \le \frac{2}{n}$.
\item If $i_1=i_2$ (or symmetrically $i_3=i_4$) then 
\[
\cov{W_{B,i_1}}{W_{B,i_2} W_{B,i_3}} = \expe{W_{B,i_1} W_{B,i_2} W_{B,i_3}} - \expe{W_{B,i_1}} \expe{W_{B,i_1} W_{B,i_2}} =0,
\]
by Lemma \ref{lem:moments}.
Also
\[
\cov{W_{B,i_1}^2}{W_{B,i_2} W_{B,i_3}}=\cov{1}{W_{B,i_2} W_{B,i_3}}=0.
\]
Therefore, in this case the covariance is zero.
\item If $i_1=i_3$ (or symmetrically $i_1=i_4$, $i_2=i_3$, $i_2=i_4$) then 
\begin{multline*}
\cov{W_{B,i_1} W_{B,i_2}}{W_{B,i_1} W_{B,i_4}}= \expe{ W_{B,i_2} W_{B,i_4}} - \expe{W_{B,i_1} W_{B,i_2}} \expe{W_{B,i_3} W_{B,i_4}}\\
=-\frac{1}{n_B-1}-\frac{1}{(n_B-1)^2}.    
\end{multline*}
In this case $(\bSigmawB)_{i_1,i_2} (\bSigmawB)_{i_3,i_4} =\frac{1}{(n_B-1)^2} $.
There are order of $\frac{n}{n_B} n_B^3=n n_B^2$ such cases. Hence the sum of the covariances of this case is bounded by a constant times
\[
\frac{n n_B^2}{n^2} \frac{1}{(n_B-1)^3}  \le \frac{4}{n}. 
\]
\item Suppose that $i_1,i_2, i_3, i_4$ are all different. Then,
\begin{multline*}
\cov{W_{B,i_1} W_{B,i_2}}{W_{B,i_3} W_{B,i_4}}\\
=\expe{ W_{B,i_1} W_{B,i_2} W_{B,i_3} W_{B,i_4}}-\expe{W_{B,i_1} W_{B,i_2}}\expe{W_{B,i_3} W_{B,i_4}}
\end{multline*}
By Lemma \ref{lem:moments} 
we have that
\[
\expe{W_{B,i_1} W_{B,i_2} W_{B,i_3} W_{B,i_4}} = \frac{3}{(n_B-3)(n_B-1)}.
\]
Recall that $\expe{W_{B,i_1} W_{B,i_2}}=\expe{W_{B,i_3} W_{B,i_4}} = - \frac{1}{n_B-1}$. Therefore we get that
\[
\cov{W_{B,i_1} W_{B,i_2}}{W_{B,i_3} W_{B,i_4}} = \frac{2 n_B}{(n_B-3)(n_B-1)^2}\le \frac{4}{n_B} 
\]
for $n_B \ge 4$.
In this case $(\bSigmawB)_{i_1,i_2}(\bSigmawB)_{i_3,i_4} =\frac{1}{(n_B-1)^2} $.

There are order of $\frac{n}{n_B} n_B^4=n n_B^3$ such cases. Hence the sum of the covariances of this case is bounded by a constant times
\[
\frac{n n_B^3}{n^2} \frac{1}{(n_B-1)^2}\frac{1}{n_B} \le \frac{2}{n}. 
\]
\end{itemize}
The proof of the lemma is thus concluded. \qed

\begin{lemma}\label{lem:moments}
\begin{enumerate}[(i)]
    \item Suppose that $i_1,i_2$ are different indexes and belong to the same block, then $\expe{W_{B,i_1}W_{B,i_2}}=-\frac{1}{n_B-1}$.
    \item Suppose that $i_1,i_2,i_3$ are different indexes and belong to the same block, then\\ $\expe{W_{B,i_1}W_{B,i_2}W_{B,i_3}}=0$.
    \item Suppose that $i_1,i_2,i_3,i_4$ are different indexes and belong to the same block, then\\ $\expe{W_{B,i_1}W_{B,i_2}W_{B,i_3}W_{B,i_4}}=\frac{3}{(n_B-3)(n_B-1)}$.
\end{enumerate} 
\end{lemma}

\noindent{\bf Proof of Lemma \ref{lem:moments}.} {\em Part (i).} Consider the first block and the indexes $1,2$ with out loss of generality. Let $1,\ldots,n_B$ denote the indexes of the block. We have that $\sum_{i =1}^{n_B} W_{B,i}=0$. Therefore, 
\[
0=\sum_{i=1}^{n_B} \expe{ W_{B,i} W_{B,1}}=\expe{W_{B,1}^2}+(n-1)\expe{W_{B,1}W_{B,2}} = 1+(n-1)\expe{W_{B,1}W_{B,2}},
\]
where the first equality is due to symmetry and the second is true because $W_{B,1}^2=1$. The result follows.\\
{\em Part (ii)}. Consider the indexes $1,2,3$ without loss of generality. Similar to the first part we have that
\begin{multline*}
0=\sum_{i =1}^{n_B} \expe{ W_{B,i} W_{B,1} W_{B,2}}=2 \expe{W_{B,1}^2 W_{B,2}}+(n-2)\expe{W_{B,1}W_{B,2}W_{B,3}} \\
= 0+(n-2)\expe{W_{B,1}W_{B,2}W_{B,3}}.    
\end{multline*}
It follows that $\expe{W_{B,1}W_{B,2}W_{B,3}}=0$.\\
{\em Part (iii)}. Consider the indexes $1,2,3,4$ without loss of generality. Similar to the previous parts we have that
\begin{multline*}
0=\sum_{i =1}^{n_B} \expe{ W_{B,i} W_{B,1} W_{B,2} W_{B,3}}=3 \expe{W_{B,1}^2 W_{B,2} W_{B,3}}+(n-3)\expe{W_{B,1}W_{B,2}W_{B,3}W_{B,4}} \\
= -3 \frac{1}{n_B-1}+(n-3)\expe{W_{B,1}W_{B,2}W_{B,3}W_{B,4}}.    
\end{multline*}
It follows that $\expe{W_{B,1}W_{B,2}W_{B,3}W_{B,4}}=\frac{3}{(n_B-3)(n_B-1)}$. \qed

\pagebreak
\section{More Simulation Results}

\begin{figure}[htp]
    \centering
    \includegraphics[width=\linewidth]{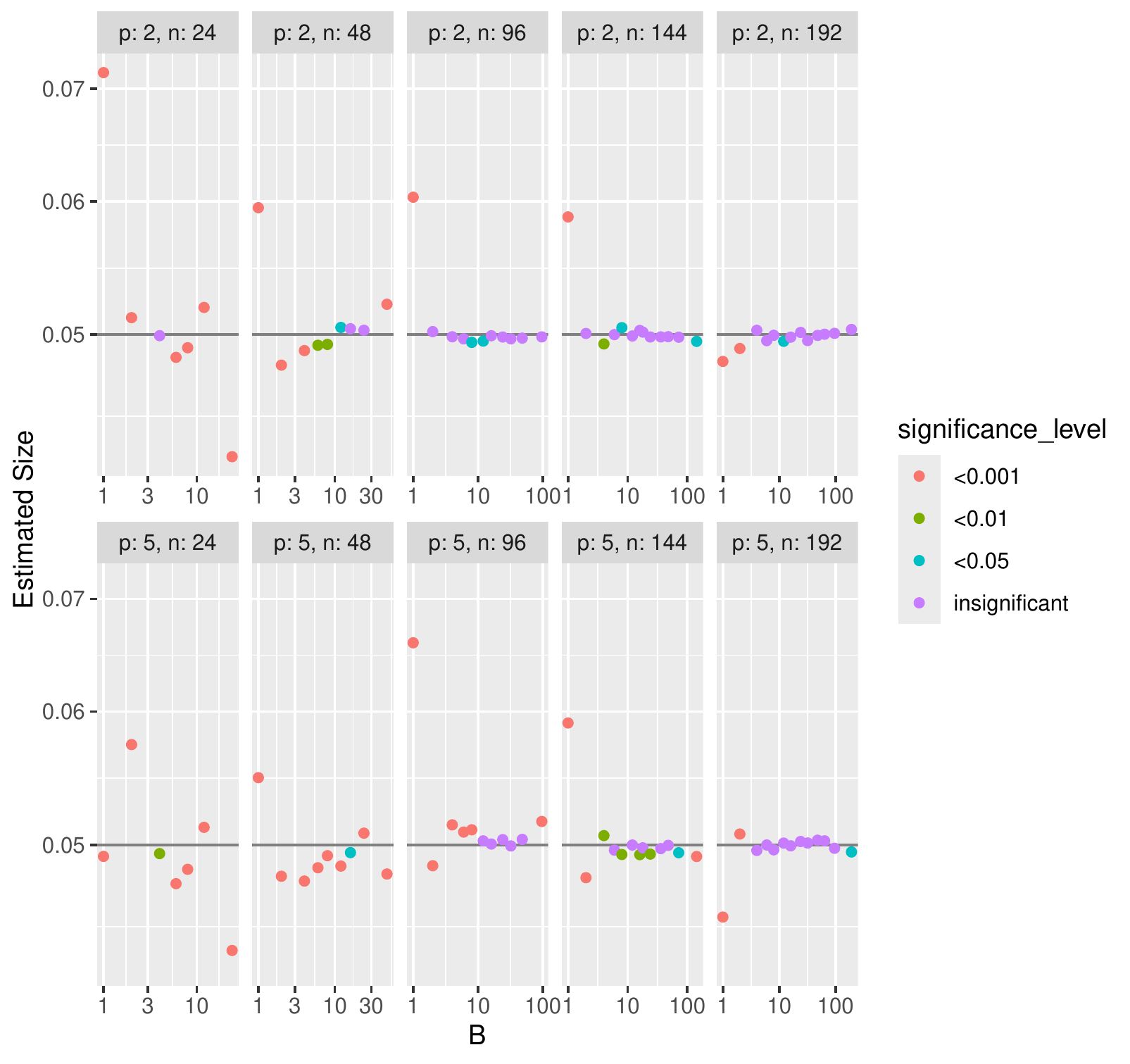}
    \caption{Size results with significance level for all 148 sample size cross block designs considered. Confidence intervals are smaller than the dots and thus unshown. \qu{Significance level} is displayed for a one-proportion two-sided z-test where $H_0: \alpha = 5\%$.}
    \label{fig:sizes_p_gr_1}
\end{figure}

\end{document}